\documentclass[3p,times,twocolumn]{elsarticle}

\usepackage{ecrc}

\volume{00}

\firstpage{1}

\journalname{Nuclear Physics B Proceedings Supplement}

\runauth{}

\jid{nuphbp}

\jnltitlelogo{Nuclear Physics B Proceedings Supplement}

\usepackage{amssymb}
\usepackage[figuresright]{rotating}

\newcommand{\gev}{\,\textrm{GeV}}

\newcommand{\eV}{\,\mathrm{eV}}
\newcommand{\fde}{f_{\rm DE}}
\newcommand{\Mp}{M_{\rm P}}
\newcommand{\ndw}{N_{\rm DW}}
\newcommand{\Mpt}{$M_{\rm P}$}
\newcommand{\Mgt}{$M_{\rm GUT}$}
\newcommand{\Mg}{M_{\rm GUT}}

\newcommand{\Ude}{U(1)$_{\rm de}$}

\newcommand{\UPQ}{U(1)$_{\rm PQ}$}

\newcommand{\ie}{{\it i.e.~}}
\newcommand{\etal}{{\it et al.}}
\begin{document}

\begin{frontmatter}



\dochead{}

\title{Dark energy, QCD axion, BICEP2, and trans-Planckian decay constant   
\tnoteref{label1}
}
\tnotetext[label1]{This work is supported by the NRF grant funded by the Korean Government (MEST) (No. 2005-0093841).}

\author{Jihn E. Kim}
\ead{jihnekim@gmail.com}

\address{Department of Physics, Kyung Hee University, Seoul 130-701, Korea}

\begin{abstract}
Discrete symmetries allowed in string compactification are the mother of all global symmetries
which are broken at some level. We discuss the resulting pseudo-Goldstone bosons, in particular the QCD axion and a temporary cosmological constant, and inflatons. We also comment on some implications of the recent BICEP2 data. 
\end{abstract}


\end{frontmatter}

% \linenumbers

\section{Discrete symmetries} \label{sec:discrete}


%% References

\nocite{*}
\bibliographystyle{elsarticle-num}
\bibliography{martin}

\begin{figure}[!b]
\centerline{\includegraphics[width=0.25\textwidth]{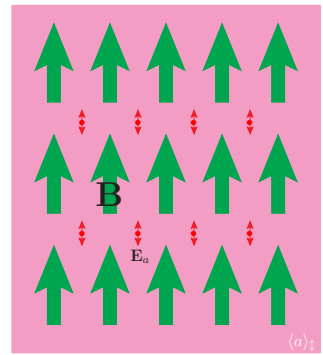}
 }
\caption{The resonant detection idea of the QCD axion. The {\bf E}-field follows the axion vacuum oscillation.}\label{Fig:AxVacuum}
\end{figure}

The cosmic energy pie is composed of 68\% dark energy (DE), 27\% cold dark matter (CDM), and 5\% atoms \cite{Planck13}. Among these, some of DE and CDM can be bosonic coherent motions (BCMs) \cite{KimYannShinji}.
The ongoing search of the QCD axion is based on the BCM. 
Being a pseudo-Goldstone boson, the QCD axion can be a composite one \cite{KimAxComp}, but after the discovery of the fundamental Brout-Englert-Higgs (BEH) boson, the possibility of the QCD axion being fundamental gained much more weight. The ongoing axion search experiment is based on the resonance enhancement of the oscillating  {\bf E}-field following the axion vacuum oscillation as depicted in Fig. \ref{Fig:AxVacuum}. It may be possible to detect the CDM axion even its contribution to CDM is only 10\% \cite{cappsite}.

 %%%%%%%%%%%%%%%%%%%%%%%%%%%%%%%%%%%%%%%%%%%%%%%%%%%%
%%%%%%%%%%%%%%%%%%%%%%%%%%%%%%%%%%%%%%%%%%%%%%%%%%%%

The BEH boson is fundamental. The QCD axion may be fundamental. The inflaton may be fundamental.  These bosons with canonical dimension 1  can affect more importantly to low energy physics compared to those of spin-$\frac12$ fermions of the canonical dimension $\frac32$.
This leads to a BEH portal to the high energy scale to the axion scale or even to the standard model (SM) singlets at the grand unification (GUT) scale. Can these singlets explain both DE and CDM in the Universe? Because the axion decay constant $f_a$ can be in the intermediate scale, axions
can live up to now ($m_a <24\, \eV$) and constitute DM of the
Universe. In this year of a GUT scale VEV,  can these also explain  the inflation finish?

 %%%%%%%%%%%%%%%%%%%%%%%%%%%%%%%%%%%%%%%%%%%%%%%%%%%%
\begin{figure}[!t]
\centerline{\includegraphics[width=0.25\textwidth]{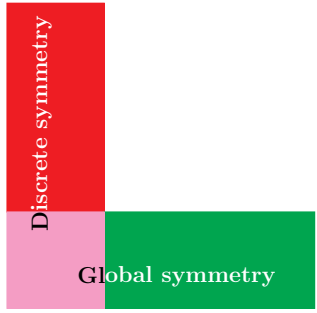}
 }
\caption{Terms respecting discrete and global symmetries.}\label{Fig:discrete}
\end{figure}
%%%%%%%%%%%%%%%%%%%%%%%%%%%%%%%%%%%%%%%%%%%%%%%%%%%%

For pseudo-Goldstone bosons like axion, we introduce global symmetries. 
 But global symmetries are known to be broken by the quantum gravity effects, especially via the Planck scale  wormholes.   To resolve this dilemma, we can think of two possibilities of discrete symmetries below \Mpt\,\cite{Kim13worm}: (i) The discrete symmetry arises as  a part  of  a gauge symmetry, 
and (ii) The string selection rules directly give the discrete symmetry. So, we will consider discrete gauge symmetries allowed in string compactification. Even though the Goldstone boson directions in spontaneously broken gauge symmetries are flat, the Goldstone boson directions of spontaneously broken {\em global} symmetries are not flat, \ie global symmetries are always {\em approximate}. The question is what is the degree of the {\em approximateness}. In Fig. \ref{Fig:discrete}, we present a cartoon separating effective terms according to string-allowed discrete symmetries. The terms in the  vertical column represent exact symmmetries such as gauge symmetries and string allowed discrete symmetries. If we consider a few terms in the lavender part, we can consider a {\em global symmetry}. With the global symmetry, we can consider the global symmetric terms which are in the lavender and green parts of Fig. \ref{Fig:discrete}. The global symmetry is broken by the terms in the red part.

The most studied global symmetry is the Peccei-Quinn (PQ) symmetry \UPQ\,\cite{PQ77}. For \UPQ, the dominant breaking term is the QCD anomaly term $(\overline{\theta}/32\pi^2)G_{\mu\nu}\tilde{G}^{\mu\nu}$ where $G_{\mu\nu}$ is the gluon field strength. Since this $\overline{\theta}$ gives a neutron EDM (nEDM) of order $10^{-16}\overline{\theta}\,e$cm, the experimental upper bound on nEDM restricts $|\overline{\theta}|<10^{-11}$. ``Why is $\overline{\theta}$ so small?" is the strong CP problem. There have been a few solutions, but the remaining plausible solution is the very light axion solution \cite{KimRMP}. In field theory, it is usually talked about in terms of the KSVZ axion \cite{KSVZ} and the DFSZ axion \cite{DFSZ}, and there are several possibilities even for these one heavy quark or one pair of BEH doublets  \cite{Kim98}. For axion detection through the idea of Fig. \ref{Fig:AxVacuum}, the axion-photon-photon coupling  $c_{a\gamma\gamma}$ is the key parameter. In our search of an ultra-violet completed theory, the customary numbers of   \cite{Kim98} are ad hoc. From string theory, so far there is only one calculation on $c_{a\gamma\gamma}$ \cite{Kimagg14}. To calculate $c_{a\gamma\gamma}$, the model must lead to acceptable SM phenomenology, otherwise the calculation does not lead to a useful global fit to all experimental data.

 %%%%%%%%%%%%%%%%%%%%%%%%%%%%%%%%%%%%%%%%%%%%%%%%%%%%
\begin{figure}[!b]
\centerline{\includegraphics[width=0.5\textwidth]{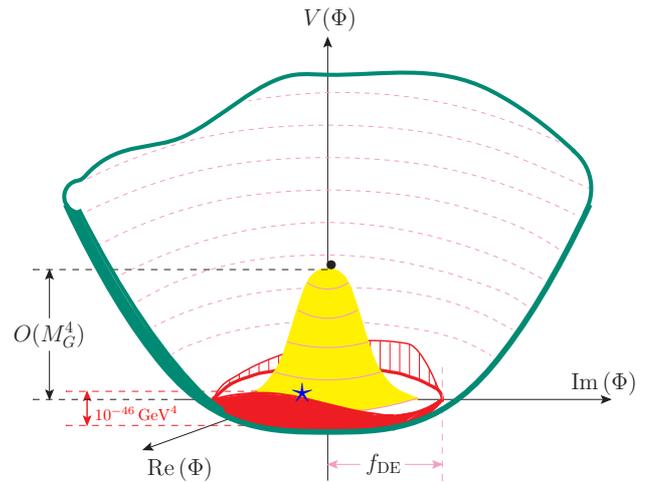}
 }
\caption{The DE potential in the red angle direction in the valley of radial field of height  $\approx \Mg^4$.}\label{Fig:HilltopV}
\end{figure}
%%%%%%%%%%%%%%%%%%%%%%%%%%%%%%%%%%%%%%%%%%%%

%%%%%%%%%%%%%%%%%%%%%%%%%%%%%%%%%%%%%
\section{Dark energy and QCD axion}
\label{sec:axionde}

It is interesting to note that the QCD axion must arise if one tries to introduce the DE scale via the idea of Fig. \ref{Fig:discrete} \cite{KimNilles14,KimJKPS14}. The DE and QCD axions are the BCM examples. Dark energy  is classified as {\bf CCtmp} and QCD axion  is classified as {\bf BCM1} in \cite{KimYannShinji}. 

Note that the global symmetry violating terms belong to the red part in Fig. \ref{Fig:discrete}. For the QCD axion, the dominant breaking is by the QCD anomaly term, which leads to the QCD axion mass in the range of milli- to nano-eV for $f_a\simeq 10^{9-15}\,\gev$. In the BEH portal scenario, the DE pseudoscalar must couple to the color anomaly since it couples to the BEH doublet and the BEH scalar couples to the quarks. On the other hand, a {\bf CCtmp} psudoscalar mass is in the range $10^{-33}\sim 10^{-32}\,\eV$ \cite{Carroll98}. Therefore, the QCD anomaly term is too large to account for the DE scale of $10^{-46\,}\gev^4$, and we must find out a QCD-anomaly free global symmetry. It is possible by introducing two global U(1) symmetries  \cite{KimNilles14,KimJKPS14}. 

In addition, the breaking scale of \Ude~is trans-Planckian \cite{Carroll98}. 
Including the anharmonic term carefully with the new data on light quark masses, a recent calculation of the cosmic axion density gives the axion window \cite{Bae08},
\begin{equation} 
10^{9\,}\gev<f_a<10^{12\,}\gev.\label{eq:window}
\end{equation}
 It is known that string axions from $B_{MN}$ have GUT scale decay constants \cite{ChoiKimfa85}; hence the QCD axion from string theory is better to arise from matter fields  \cite{KimPRL14}. For the QCDaxion, the height of the potential is $\approx\Lambda_{\rm QCD}^4$. For the DE pseudo-Goldstone boson, the height of the potential is $\approx\Mg^4$, according to the BEH portal idea, as shown in 
Fig. \ref{Fig:HilltopV}. With \UPQ\,and \Ude, one can construct a DE model from string compactification \cite{KimJKPS14}. Using the SUSY language,
the discrete and global symmetries below $\Mp$ are the consequence of the full superpotential $W$. So, the exact symmetries related to string compactification are respected by the full $W$, \ie the vertical column of Fig. \ref{Fig:discrete}. Considering only the $d=3$ superpotential $W_3$,  we can consider an
approximate PQ symmetry.
For the MSSM interactions supplied by R-parity, one needs to know all the SM singlet spectrum. We need ${\bf Z}_2$ for a WIMP candidate. 

 Introducing  two global symmetries, we can remove the \Ude-$G$-$G$ where $G$ is QCD and the \Ude~charge is a linear combination of two global symmetry charges. The decay constant corresponding to \Ude~is $\fde$.  Introduction of two global symmetries is inevitable to interpret the DE scale and hence in this scenario the appearance of \UPQ~is a natural consequence. The height of DE potential  is so small, $10^{-46\,}\gev^4$, that the needed discrete symmetry breaking term of Fig. \ref{Fig:discrete} must be small, implying the discrete symmetry is of high order. Now, We have a scheme to explain both 68\% of DE and 27\% of CDM via approximate {\em global} symmetries. With SUSY, axino may contribute to CDM also \cite{Baer14}.

A typical example for the  discrete symmetry is ${\bf Z}_{10\,R}$ as shown in \cite{KimJKPS14}. The ${\bf Z}_{10\,R}$  charges descend from a gauge U(1) charges of the string compactification \cite{HuhKK09}. Then, the height of the potential is highly suppressed and we can obtain $10^{-47\,}\gev^4$, without the gravity spoil of the global symmetry. In this scheme with BEH portal, we introduced three  scales for vacuum expectation values (VEVs), TeV scale for $H_uH_d$, the GUT scale \Mgt~for singlet VEVs, and the intermediate scale for the QCD axion. The other fundamental scale is $\Mp$. The trans-Planckian decay constant $\fde$ can be a derived scale \cite{KNP05}.

Spontaneous breaking of \Ude~is via a Mexican hat potential with the height of $\Mg^4$.
A byproduct of this Mexican hat potential is the hilltop inflation with the height of $O(\Mg^4)$,
as shown in Fig. \ref{Fig:HilltopV}.
It is a small field inflation, consistent with the 2013 Planck data.

\section{Gravity waves  from U(1)$_{\rm de}$ potential}
\label{sec:gravitywave}
%%%%%%%%%%%%%%%%%%%%%%%%%%%%%%%%%%%%%%%%%%%%%%%%%%%%%%%%%%%%%%%%%%%%%%%%%%%
 
 However,   with the surprising report from the BICEP2 group on a large tensor-to-scalar ratio $r$  \cite{BICEP2}, we must reconsider the above hilltop inflation whether it leads to appropriate numbers on $n_s, r$ and the e-fold number $e$, or not. With two U(1)'s, the large trans-Planckian $\fde$ is not spoiled by the intermediate PQ scale $f_a$ because the PQ scale just adds to the $\fde$ decay constant only by a tiny amount, viz. $\fde\to \sqrt{\fde^2+O(1)\times f_a^2} \approx\fde$ for $|f_a /\fde\simeq 10^{-7}|$.  

Inflaton potentials with almost flat one near the origin, such as the Coleman-Weinberg type new inflation, were the early attempts for inflation. But any models can lead to inflation if the potential is flat enough as in the {\em chaotic inflation} with small parameters \cite{Linde83}. A single field chaotic inflation survived until now is the $m^2\phi^2$ scenario {\em chaotic inflation} with $m=O(10^{13\,}\gev)$. To shrink the field energy much lower than $\Mp^4$, a {\em natural inflation} (mimicking the axion-type $-\cos$ potential) has been introduced \cite{Freese90}. If a large $r$ is observed, Lyth noted that the field value $\langle\phi\rangle$ must be larger than $15\,\Mp$, which is known as the Lyth bound \cite{Lyth97}. To obtain this trans-Planckian field value, the Kim-Nilles-Peloso (KNP) 2-flation has been introduced with two axions \cite{KNP05}. It is known recently that the natural inflation is more than $2\sigma$ away from the central value of BICEP2, $(r,n_s)=(0.2,0.96)$. In general, the hilltop inflation  gives almost zero $r$. This is because $n_s\simeq 1-\frac38 r+2\eta$ which gives $n_s=0.925$ for $(r,\eta)=(0.2,0)$. To raise $n_s$ from $0.925$ to $0.96$, we need a positive $\eta$, but the hilltop point gives a negative $\eta$. 
 
Therefore, for the \Ude\,hilltop inflation to give a suitable $n_s$ with a large $r$, one must introduce another field which is called {\em chaoton} because it provides the behavior of $m^2\phi^2$ term at the BICEP2 point \cite{KimHilltop14}. With this hilltop potential, the height is of order $\Mg^4$ and the decay constant is required to be $> 15\,\Mp$. Certainly, the potential energy is smaller than order $\Mp^4$  for
$\phi=[0,\fde]$. Since this hilltop potential is obtained from the mother discrete symmetry, such as ${\bf Z}_{10\,R}$, the flat valley up to the trans-Planckian $\fde$ is possible, for which the necessary condition is given in terms of quantum numbers of  ${\bf Z}_{10\,R}$ \cite{KimHilltop14}.
 
 %%%%%%%%%%%%%%%%%%%%%%%%%%%%%%%%%%%%%%%%%%%%%%%%%%%%
\begin{figure}[!b]
\centerline{\includegraphics[width=0.38\textwidth]{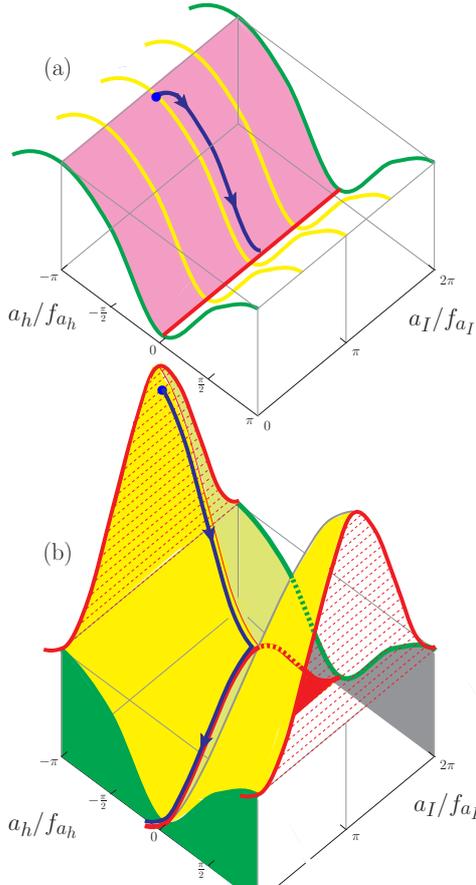}   }
\caption{Two-flation. (a) The flat valley with one confining force, and (b) the KNP model with two confining forces.}\label{Fig:twofl}
\end{figure}
%%%%%%%%%%%%%%%%%%%%%%%%%%%%%%%%%%%%%%%%%%%%

%%%%%%%%%%%%%%%%%%%%%%%%%%%%%%%%%%%%%
\section{The KNP model and \Ude~hilltop inflation}
\label{sec:KNPtrans}

A large VEV of a scalar field is possible if a very small coupling constant $\lambda$ is assumed in $V=\frac14\lambda(|\phi|^2-f^2)^2 $ with a small mass parameter $m^2=\lambda f^2$. With a GUT scale $m$, $f$ can be trans-Planckian of order $10\Mp$ for $\lambda< 10^{-6}$. But this potential is a single field hilltop type and it is not favored by the above argument with the BICEP2 data \cite{KimHilltop14}. This has led to the recent surge of studies on concave potentials near the origin of the single field. The concave potentials give positive $\eta$'s.

To cut  off the potential exceeding the GUT scale $\Mg^4$, the natural inflation with a GUT scale confining force has been introduced \cite{Freese90}. With two confining forces, it was
possible to raise a decay constant of the GUT scale axions above \Mpt, which is known as the Kim-Nilles-Peloso (KNP) 2-flation model \cite{KNP05}. In terms of
two axions $a_1$ and $a_2$ and two GUT scale ($\Lambda_1$ and $\Lambda_2$) confining forces, the minus-cosine potentials can be written as
\begin{eqnarray}
V&=&\Lambda_{1}^4\left(1-\cos\left[\alpha\frac{a_1}{f_1}+\beta\frac{a_2}{f_2})\right]\right)
\nonumber\\
&+& \Lambda_{2}^4\left(1-\cos\left[\gamma\frac{a_1}{f_1}+\delta\frac{a_2}{f_2})\right]\right) \,,\label{eq:KNP2}
\end{eqnarray}
where $\alpha,\beta,\gamma$, and $\delta$ are determined by two U(1) quantum numbers.
If there is only one confining force, we can set $\Lambda_2=0$ in Eq. (\ref{eq:KNP2}), which is depicted in Fig. \ref{Fig:twofl}\,(a). The flat red valley cannot support the inflation energy.
The situation with two confining forces is shown in Fig. \ref{Fig:twofl}\,(b). The inflation path is shown as the arrowed blue curve on top of the red valley on the yellow roof. 
In this case, we consider a $2\times 2$ mass matrix,
\begin{eqnarray}
M^2=\left(\begin{array}{cc}
   \frac{1}{f_1^2}\left({\alpha^2\Lambda^4_1}+ {\gamma^2\Lambda^4_2}
   \right), &\frac{1}{f_1f_2} (\alpha\beta\Lambda^4_1+ \gamma\delta\Lambda^4_2) 
   \\[1em]
   \frac{1}{f_1f_2}(\alpha\beta\Lambda^4_1+ \gamma\delta\Lambda^4_2) , &   
 \frac{1}{f_2^2}(\beta^2\Lambda^4_1+\delta^2\Lambda^4_2) \end{array}
\right) \nonumber
 \end{eqnarray}
whose eigenvalues are \cite{ChoiKY14}, $m_h^2= \frac12(A+B)$ and 
 \begin{eqnarray} 
m_I^2&=&\frac12(A-B)\,. \label{eq:massMI}
 \end{eqnarray}
 with
 \begin{eqnarray} 
A&=&\left(\frac{\alpha^2\Lambda_1^4 
+\gamma^2\Lambda_2^4}{f_1^2} +\frac{\beta^2\Lambda_1^4 
+\delta^2\Lambda_2^4}{f_2^2}\right),\nonumber\\ 
B&=&\sqrt{A^2 -4(\alpha\delta -\beta\gamma)^2 \frac{\Lambda_1^4 \Lambda_2^4}{f_1^2 f_2^2} }\,. \label{eq:massMI}
 \end{eqnarray}
From Eq. (\ref{eq:massMI}), we note that a large $f_I$ is possible for $\alpha\delta= \beta\gamma+\Delta$ with $\Delta\approx 0$.  
Then, the inflaton mass is
\begin{eqnarray}
m_I^2 &\simeq& \frac{\Delta^2\Lambda_1^4\Lambda_2^4}{f_2^2(\alpha^2\Lambda_ 1^4 +\gamma^2\Lambda_2^4) +f_1^2(\beta^2\Lambda_1^4 
+\delta^2\Lambda_2^4)}\,.\nonumber
  \end{eqnarray}
For $\Lambda_1=\Lambda_2$ and $f_1=f_2\equiv f$, it becomes
\begin{eqnarray}
m_I^2&\simeq& \frac{\Lambda^4}{(\alpha^2 +\beta^2 +\gamma^2+\delta^2 )f^2/\Delta^2}\,.
\label{eq:mIapp} 
 \end{eqnarray}
The PQ quantum numbers $\alpha,\beta,\gamma$, and $\delta$ are not random priors, but given definitely in a specific model. The perspectives of 2- and N-flations are given in \cite{ChoiKY14}. 

Even though a large trans-Planckian decay constant is in principle possible with large PQ quantum numbers in the KNP model, string compactification may not allow that possibility. The N-flation with a large N has a more severe problem in string compactification \cite{ChoiKY14}. 
This invites to look for another possibility of generating trans-Planckian decay constants.  Since the KNP model already introduced two axions, we look for a possibility of introducing another field (called chaoton before) in the hilltop potential. In effect, the chaoton is designed to provide a positive $\eta$.

\begin{figure}[!t]
\centerline{\includegraphics[width=0.45\textwidth]{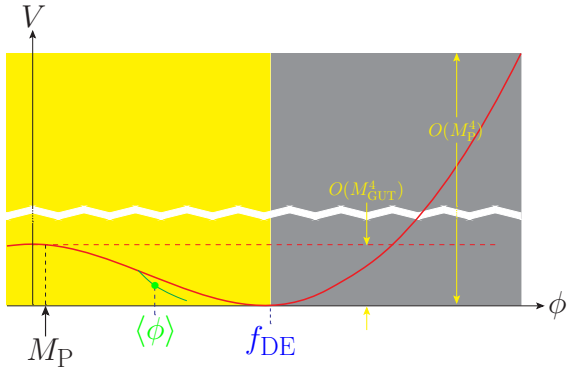}
 }
\caption{The trans-Planckian decay costant in the hilltop inflation.}\label{Fig:Trans}
\end{figure}

The hilltop potential of Fig. \ref{Fig:HilltopV} is a Mexican hat potential of \Ude, \ie obtained from some discrete symmetry, allowed in string compactification \cite{KimNilles14}. The discrete symmetry may provide a small DE scale. The trans-Planckian decay constant, satisfying the Lyth bound, is obtained by a small coupling $\lambda$ in the hilltop potential $V$. The requirement for the vacuum energy being much smaller than $\Mp^4$ is achieved by restricting the inflaton path in the hilltop region, $\langle\phi \rangle \lesssim \fde$, as shown in Fig. \ref{Fig:Trans}. In Fig. \ref{Fig:Trans}, the inflation path affected by chaoton is depicted as the green path.

 %%%%%%%%%%%%%%%%%%%%%%%%%%%%%%%%%%%%%%%%%%%%%%%%%%%%
%%%%%%%%%%%%%%%%%%%%%%%%%%%%%%%%%%%%%%%%%

We can compare this hilltop inflation assisted by chaoton with the $m^2\phi^2$ chaotic inflation. The hilltop inflation is basically a consequence of discrete symmetries \cite{Kim13worm,KimNilles14,KimJKPS14} , allowed in string compactification. If some conditions are satisfied between the discrete quantum numbers of the GUT scale fields and trans-Planckian scale fields, the hilltop potential of Fig.  \ref{Fig:Trans} can result \cite{KimHilltop14}. On the other hand, the $m^2\phi^2$ chaotic inflation does not have such symmetry argument, and lacks a rationale forbidding higher order $\phi^n$ terms. This argument was used to forbid many interesting theories by considering the observed slow-roll parameter $\eta$ from inflation assumption \cite{Lyth14}. But the situation is much worse
here than Lyth's case. For example, for an $n=104$ term for the trans-Planckian field $\Phi$ and the GUT scale field $\phi$, one must fine-tune the coupling 1 out of $10^{127}$ for the trans-Planckian singlet VEV of order $\langle\Phi \rangle\approx 31\Mp$ \cite{KimHilltop14}.

%%%%%%%%%%%%%%%%%%%%%%%%%%%%%%%%%%%
\section{PQ symmetry breaking below $H_I$} \label{sec:PQscale}

Cosmology of axion models was  started in 1982--1983 \cite{Preskill83} with the micro-eV axions \cite{KSVZ,DFSZ}. The needed axion scale given in Eq. (\ref{eq:window}), far below the GUT scale, is understood in models with the anomalous U(1) in string compactification \cite{Kim88}. In addition to the scale problem, there exists the cosmic-string and domain wall (DW) problem \cite{Vilenkin82,Sikivie82}. Here, I want to stress that the axion DW problem
has to be resolved without the dilution effect by inflation. 

 %%%%%%%%%%%%%%%%%%%%%%%%%%%%%%%%%%%%%%%%%%%%%%%%%%%%
\begin{figure}[!t]
\centerline{\includegraphics[width=0.3\textwidth]{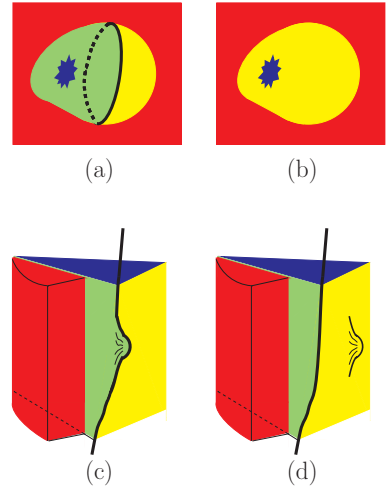} }
\caption{Small DW balls ((a) and (b), with punches dshowing the inside blue-vacuum) and the horizon scale string-wall system ((c) and (d))  for $\ndw=2$. Yellow walls are $\theta=0$ walls, and  yellow-green walls are $\theta=\pi$ walls.  Yellow-green walls of type (b) are also present.}\label{Fig:DWtwo}
\end{figure}
%%%%%%%%%%%%%%%%%%%%%%%%%%%%%%%%%%%%%%%%%%%%

\begin{figure}[!b]
\centerline{\includegraphics[width=0.38\textwidth]{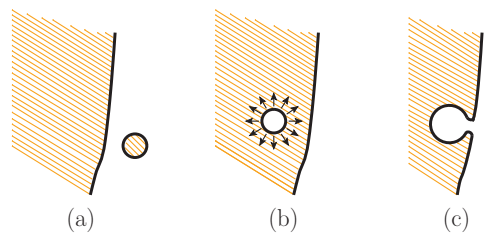} }
\caption{The horizon scale string-wall system with $\ndw=1$. Any point is connected to another point, not passing through the wall. .}\label{Fig:DWone}
\end{figure}

The BICEP2 finding of ``high scale inflation at the GUT scale'' implies the reheating temperature after inflation $\gtrsim 10^{12\,}\gev$. Then, studies on the isocurvature constraint with the BICEP2 data
pin down the axion mass in the upper allowed region \cite{Marsh14}.  But this axion mass is based on the numerical study of Ref. \cite{Kawasaki12} which has not included the effects of axion string-DW annihilation by the Vilenkin-Everett mechanism \cite{Vilenkin82}.   In Fig. \ref{Fig:DWtwo}, we present the case for $\ndw=2$. Topological defects are small balls ((a) and (b)), whose walls separarte $\theta=0$ and $\theta=\pi$ vacua, and a horizon scale string-wall system. Collisions of small balls on the horizon scale walls do not punch a hole, and the horizon size string-DW system is not erased ((c) and (d)). Therefore, for $\ndw\ge 2$ axion models, there exists the cosmic energy crisis problem of the string-DW system.   In Fig. \ref{Fig:DWone}, we present the case with $\ndw=1$. Topological defects are small disks and a horizon scale string-DW system ((a)). Collisions of small balls on the horizon scale walls punch holes ((b)), and the holes expand with light velocity. In this way, the string-wall system is erased  ((c)) and the cosmic energy crisis problem is not present in $\ndw=1$ axion models \cite{BarrChoiKim}, for example with one heavy quark in the KSVZ model. If the horizon-scale string-DW system is absent, there is no severe axion DW problem. 

 %%%%%%%%%%%%%%%%%%%%%%%%%%%%%%%%%%%%%%%%%%%%%%%%%%%%
%%%%%%%%%%%%%%%%%%%%%%%%%%%%%%%%%%%%%%%%%%%%

So, with the BICEP2 report, it became of utmost importance to obtain $\ndw=1$ axion models.
The first try along this line was the so-called Lazarides-Shafi mechanism, using the center (discrete group) of GUT gauge groups \cite{LS82}. A more useful discrete group is a discrete subgroup of continuous U(1)'s, \ie the discrete points of the longitudinal Goldstone boson directions of gauged U(1)'s \cite{ChoiKimDW85}. In string theory, the anomalous gauged U(1) is useful for this purpose \cite{Kim88}. This solution has been recently obtained in ${\bf Z}_{12-I}$ orbifold compactification \cite{KimDW14}.

 %%%%%%%%%%%%%%%%%%%%%%%%%%%%%%%%%%%%%%%%%%%%%%%%%%%%
\begin{figure}[!t]
\centerline{\includegraphics[width=0.43\textwidth]{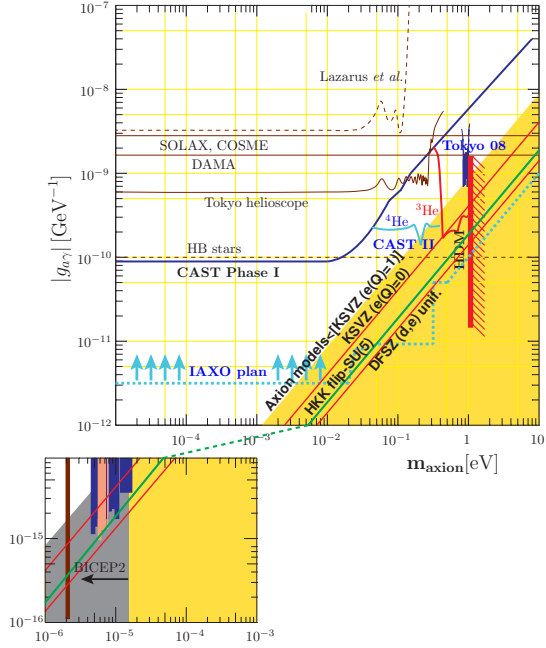} }
\caption{The $g_{a\gamma}(=1.57\times 10^{-10\,}c_{a\gamma\gamma})\,$ vs. $m_a$ plot \cite{Baerprp}.}\label{Fig:expbound}
\end{figure}
%%%%%%%%%%%%%%%%%%%%%%%%%%%%%%%%%%%%%%%%%%%%

The QCD-axion string-DW problem may not appear at all if the hidden-sector confining gauge theory conspire to erase the hidden-sector string-DW system \cite{BarrKim14}. Here, we introduce just one axion, namely through the anomalous U(1) gauge group, surviving down to the axion window as a global \UPQ. Here, we introduce two kinds of heavy quarks, one the SU($N_h$) heavy quark $Q_h$ and the other SU(3)$_{\rm QCD}$ heavy quark $q$. Then, the type of Fig. \ref{Fig:DWone} is present with two kinds of walls: one of $\Lambda_{h}$ wall and the other of $\Lambda_{\rm QCD}$ wall. But, at $T\approx \Lambda_{h}$ only $\Lambda_{h}$ wall is attached. At somewaht lower temperature $T_{\rm er}$ ($< \Lambda_{h}$) the string-DW system is erased {\it \`a la} Fig. \ref{Fig:DWone}. The height of the  $\Lambda_{h}$ wall is proportional to $m_{Q_h}\Lambda_h^3$ with $m_{Q_h}=f\langle X\rangle $. The VEV $\langle X\rangle $ is temperature dependent, and it is possible that  $\langle X\rangle =0$ below some critical temperature $T_c\,(<T_{\rm er})$. Then, the  $\Lambda_{h}$ wall  is erased below $T_c$, and at the QCD phase transition only the QCD wall is present. But, all horizon scale strings have been erased already and there is no energy crisis problem of the QCD-axion string-DW system. Therefore, pinpointing the axion mass using the numerical study of Ref. \cite{Kawasaki12} is not water-proof.

The $\ndw=1$ models are very attractive and it has been argued that the model-independent axion in string models, surviving down as a \UPQ\,symmetry below the anomalous U(1) gauge boson mass scale, is good for this. At the intermediate mass scale $Q_{\rm PQ}=1$ should obain a VEV to have  $\ndw=1$. In a ${\bf Z}_{12-I}$ orbifold compactification present in Ref. \cite{HuhKK09}, the axion-photon-photon coupling has been calculated \cite{Kimagg14},
\begin{equation} 
c_{a\gamma\gamma}=\frac{1123}{388} -0.98\simeq 0.91,
\end{equation}  
which is shown as the green line in the axion coupling vs. axion mass plot, Fig. \ref{Fig:expbound}. 

%%%%%%%%%%%%%%%%%%%%%%%%%%%%%%%%%%%%%%%%%%%%%%%%%%%%%%%%%%%%%%%%%%%%%%%%%%%%%%%%%%%%%%%%%%%%%%%%

\end{document}